\documentclass[aps,prd,10pt,nofootinbib,eqsecnum]{revtex4-2}
\usepackage[T1]{fontenc}
\usepackage[utf8]{inputenc}
\usepackage[a4paper]{geometry}
\usepackage{xcolor}
\usepackage{amsmath}
\usepackage{amssymb}
\usepackage{bm}
\usepackage{mathtools}
\usepackage{datetime2}
\usepackage[unicode=true,pdfusetitle,
 bookmarks=true,bookmarksnumbered=false,bookmarksopen=false,
 breaklinks=false,pdfborder={0 0 1},backref=false,colorlinks=true]
 {hyperref}



\begin{document}

\title{Polarized gravitational waves in the parity violating scalar-nonmetricity theory}

\author{Zheng Chen}%
	\affiliation{%
	School of Physics and Astronomy, Sun Yat-sen University, Zhuhai 519082, China}

\author{Yang Yu}%
	\affiliation{%
	School of Physics and Astronomy, Sun Yat-sen University, Zhuhai 519082, China}

\author{Xian Gao}%
	\email[Corresponding author: ]{gaoxian@mail.sysu.edu.cn}
	\affiliation{%
	School of Physics and Astronomy, Sun Yat-sen University, Zhuhai 519082, China}

\date{December 30, 2022}

\begin{abstract}
	There has been increasing interest in investigating the possible parity violating features in the gravity theory and on the cosmological scales. In this work, we consider a class of scalar-nonmetricity theory, of which the Lagrangian is polynomial built of the nonmetricity tensor and a scalar field. The nonmetricity tensor is coupled with the scalar field through its first order derivative. Besides the monomials that are quadratic order in the nonmetricity tensor, we also construct monomials that are cubic order in the nonmetricity tensor in both the parity preserving and violating cases. These monomials act as the non-canonical (i.e., non-quadratic) kinetic terms for the spacetime metric, and will change the behavior in the propagation of the gravitational waves. We find that the gravitational waves are generally polarized, which present both the amplitude and velocity birefringence  features due to the parity violation of the theory. Due to the term proportional to $1/k$ in the phase velocities, one of the two polarization modes suffers from the gradient instability on large scales.
\end{abstract}

\maketitle

\section{Introduction}

In light of the detection of the gravitational waves (GWs) events \cite{LIGOScientific:2016aoc,LIGOScientific:2017vwq} as well as the current and the forthcoming GWs experiments, including LISA \cite{LISA:2017pwj}, BBO \cite{Harry:2006fi}, KAGRA \cite{Kawamura:2011zz}, ET \cite{Sathyaprakash:2012jk}, Taiji \cite{Hu:2017mde,Ruan:2018tsw}, TianQin \cite{TianQin:2015yph} and Ali \cite{Li:2017drr}, the GWs have supplied us a new tool to explore the nature of gravity.
The General Relativity (GR) propagates two polarization modes of the GWs, with exactly the same amplitude and propagating speed as that of the light.
One of the interests concerning the GWs is the possible parity violation in the gravity theory as well as in the early universe, which indicates different behavior of the two polarization modes of the GWs.

One of the parity violating (PV) theories is the Chern-Simons (CS) modified gravity \cite{Lue:1998mq,Jackiw:2003pm}, which has been studied extensively in cosmology and GWs \cite{Satoh:2007gn,Saito:2007kt,Satoh:2007gn,Alexander:2009tp,Yunes:2010yf,Gluscevic:2010vv,Yagi:2012ya,Dyda:2012rj,Myung:2014jha,Alexander:2017jmt,Yagi:2017zhb,Kawai:2017kqt,Bartolo:2017szm,Bartolo:2018elp,Nair:2019iur,Nishizawa:2018srh,Odintsov:2019mlf,Fu:2020tlw,Fronimos:2021czc,Odintsov:2022hxu,Odintsov:2022cbm,Li:2022grj,Cai:2022lec,Peng:2022ttg}.
Recently, the induced GWs from the scalar perturbations in CS modified gravity has also been studied \cite{Zhang:2022xmm}. 
The CS modified gravity can be extended by including higher order derivatives of the scalar field \cite{Crisostomi:2017ugk}, which is proved to be ghostfree on a cosmological background.
While on the cosmological background or generally when the scalar field is timelike, the scalar-tensor theory is equivalent to a metric theory respecting only the spatial covariance, which we refer to as the spatially covariant gravity (SCG) \cite{Gao:2014soa,Gao:2014fra,Gao:2020yzr,Hu:2021bbo,Hu:2021yaq}.
The well-studied effective field theory of inflation \cite{Creminelli:2006xe,Cheung:2007st} and as well as the Ho\v{r}ava gravity \cite{Horava:2009uw,Blas:2009qj} can be viewed as subclasses of the SCG theories.

The Ho\v{r}ava gravity with parity violation was explored in \cite{Takahashi:2009wc,Wang:2012fi,Zhu:2013fja}.
The polarized GWs in such Lorentz breaking PV gravity models have been studied in \cite{Myung:2009ug,Cannone:2015rra,Zhao:2019szi,Zhao:2019xmm,Qiao:2019hkz,Qiao:2019wsh,Qiao:2021fwi,Gong:2021jgg}.
Within the framework of SCG, the general equations of motion for the polarized gravitational waves on a cosmological background was derived in \cite{Gao:2019liu}.
The polarized GWs exhibit interesting features such as the velocity and amplitude birefringence phenomena, i.e., the propagating velocities and the frictional terms in the equations of motion for the two polarized modes of GWs become different \cite{Alexander:2004wk,Mylova:2019jrj,Biagetti:2020lpx,Wang:2020pgu,Wang:2021gqm,Wang:2020cub,Hu:2020rub,Bartolo:2020gsh,Orlando:2022rih,Chen:2022soq,Zhao:2022pun}. 
See \cite{Zhu:2022dfq,Qiao:2022mln} for recent reviews and more references therein.

Recently there also arises interest on modified gravity theories based on non-Riemannian geometry, i.e., with torsion and/or nonmetricity tensors.
In particular, with nonmetricity tensor $Q_{\rho\mu\nu}\equiv \nabla_{\rho}g_{\mu\nu}$ and vanishing curvature tensor, symmetric teleparallel gravity and its extensions (e.g., $f(Q)$ gravity) have also been studied \cite{Nester:1998mp,Adak:2005cd,Adak:2006rx,Adak:2008gd,Mol:2014ooa,Lu:2019hra,Xu:2020yeg,BeltranJimenez:2018vdo,BeltranJimenez:2019tme,Lu:2019hra,Lazkoz:2019sjl,Albuquerque:2022eac,Dimakis:2022rkd,Zhao:2021zab,Jimenez:2022uvo}.
Through the so-called ``geometric trinity'' \citep{BeltranJimenez:2017tkd,BeltranJimenez:2019odq,Jimenez:2019woj,BeltranJimenez:2019tme,Gomes:2022vrc}, it can be shown that the curvature, torsion and nonmetricity tensors are three equivalent and complimentary approaches to describing gravity.
The ``scalar-torsion'' and ``scalar-nonmetricity'' theories, i.e., general couplings between the scalar field and torsion and/or nonmetricity tensor have also been considered in \citep{Bahamonde:2017wwk,Bahamonde:2019shr,Runkla:2018xrv,Jarv:2018bgs,Runkla:2018xrv,Hohmann:2018wxu,Hohmann:2018xnb,Soudi:2018dhv}.
See \cite{Hehl:1994ue,Heisenberg:2018vsk,Krssak:2018ywd,Bahamonde:2021gfp,Lu:2021wif} for reviews and more references therein.

The simplest term corresponding to the CS term in the presence of torsion is the so-called Nieh-Yan (NY) term \cite{Nieh:1981ww}. 
The polarized GWs have been extensively studied with NY term and its extensions (i.e., the parity violating extension of teleparallel equivalent General Relativity) \cite{Chatzistavrakidis:2020wum,Cai:2021uup,Wu:2021ndf,Langvik:2020nrs,Li:2020xjt,Li:2021wij,Rao:2021azn,Li:2022mti}, as well as in more general models with non-vanishing torsion and/or nonmetricity tensors \cite{Hohmann:2020dgy,Bombacigno:2021bpk,Iosifidis:2020dck,Hohmann:2022wrk,Conroy:2019ibo,Iosifidis:2021bad,Pagani:2015ema,Boudet:2022nub,Bombacigno:2022naf,Li:2021mdp,Li:2022vtn,Iosifidis:2018zwo}.

In this work we investigate the polarized gravitational waves in a class of parity violating scalar-nonmetricity gravity theories.
We concentrate on scalar monomials built of the nonmetricity tensor and a scalar field.
The nonmetricity tensor can be coupled to the first derivative of the scalar field.
The authors of \citep{Conroy:2019ibo} considered 3 types of PV scalar-nonmetricity monomials: $\epsilon\phi\phi QQ$, $\epsilon\phi Q\nabla Q$ and $\epsilon\phi\phi\nabla Q\nabla Q$, where $\epsilon$, $Q$ and $\phi$ stand for the Levi-Civita tensor, the nonmetricity tensor and the first order derivative of the scalar field, respectively.
Generally, monomials of the form $\sim \nabla Q\nabla Q$ are quadratic in the second order derivative of the metric and thus possibly suffer from the ghost problem.
According to the order of derivatives, monomials that are cubic in the nonmetricity tensor, i.e., in the form $\sim QQQ$ and $\sim \epsilon QQQ$, have the same importance as those in the form $\sim Q\nabla Q$.
It is thus natural to build monomials cubic order in the nonmetricity tensor and to investigate their implications on the GWs.

It is well-known that in the usual scalar-tensor theory, a non-canonical (i.e., non-quadratic) kinetic term for the scalar field will modify the propagating speed of the scalar perturbation.
In our case, the monomials in the form $\sim QQQ$ just play the same role as the non-canonical kinetic term for the spacetime metric.
One would expect that the propagating speed of the GWs will also get modified.
Therefore the observations on the propagating speed of the GWs can be used to constrain the theory.
This paper is devoted to this issue.

This paper is organized as follows. In Sec. \ref{sec:pvsq}, we build the scalar monomials in both the parity preserving and violating cases up to the cubic order in the nonmetricity tensor.
In Sec. \ref{sec:gws}, we consider the linear tensor perturbations in our model on a cosmological background, and derive the equations of motion for the gravitational waves.
In Sec. \ref{sec:con} we summarize our result.
Throughout this paper we choose the unit $8\pi G =1$ and the convention for the metric $\{-,+,+,+\}$.

\section{Parity violating scalar-nonmetricity theory} \label{sec:pvsq}

The nonmetricity tensor is defined by
	\begin{equation}
		Q_{\rho\mu\nu} \coloneqq \nabla_{\rho} g_{\mu\nu},
	\end{equation}
where $g_{\mu\nu}$ is the spacetime metric and $\nabla$ is a general affine connection.
For later convenience, we denote
	\begin{equation}
		Q_{\mu}\equiv Q_{\mu\phantom{\rho}\rho}^{\phantom{\mu}\rho} = g^{\rho\sigma} Q_{\mu\rho\sigma},\quad q_{\mu}\equiv Q_{\phantom{\rho}\rho\mu}^{\rho} = g^{\rho\sigma} Q_{\rho\sigma\mu},
	\end{equation}
for shorthands.
As in the symmetric teleparallelism, we assume that the affine connection is free of curvature and torsion,
	\begin{equation}
	R_{\phantom{\mu}\nu\rho\sigma}^{\mu} \equiv \partial_{\rho}\Gamma_{\phantom{\mu}\nu\sigma}^{\mu}-\partial_{\sigma}\Gamma_{\phantom{\mu}\nu\rho}^{\mu}+\Gamma_{\phantom{\mu}\lambda\rho}^{\mu}\Gamma_{\phantom{\lambda}\nu\sigma}^{\lambda}-\Gamma_{\phantom{\mu}\lambda\sigma}^{\mu}\Gamma_{\phantom{\lambda}\nu\rho}^{\lambda} = 0,
	\end{equation}
and
	\begin{equation}
		T_{\phantom{\rho}\mu\nu}^{\rho} \equiv \Gamma_{\phantom{\rho}\nu\mu}^{\rho}-\Gamma_{\phantom{\rho}\mu\nu}^{\rho}=0.
	\end{equation}
As a result, the coefficients of the affine connection take the general form \cite{BeltranJimenez:2017tkd,DAmbrosio:2020nqu}
	\begin{equation}
		\Gamma_{\phantom{\alpha}\beta\mu}^{\alpha}=\frac{\partial x^{\alpha}}{\partial\xi^{a}}\frac{\partial^{2}\xi^{a}}{\partial x^{\mu}\partial x^{\beta}}, \label{Gamma}
	\end{equation}
where $\xi^{a} = \xi^{a}(x)$ with $a=0,1,2,3$ are four general scalar fields.

It is well-known that in the presence of nonmetricity tensor (with vanishing torsion tensor), the Ricci scalar is given by
	\begin{equation}
		R=\mathring{R}+Q+\mathring{\nabla}_{\mu}\left(Q^{\mu}-q^{\mu}\right),\label{RicS_dec_xpl}
	\end{equation}
where the $\mathring{R}$ and $\mathring{\nabla}$ are the Ricci scalar and the covariant derivative adapted to the metric compatible Levi-Civita connection, and $Q$ is the nonmetricity scalar defined by
	\begin{equation}
	Q\coloneqq\frac{1}{4}Q_{\rho\mu\nu}Q^{\rho\mu\nu}-\frac{1}{2}Q^{\rho\mu\nu}Q_{\mu\nu\rho}-\frac{1}{4}Q_{\mu}Q^{\mu}+\frac{1}{2}q^{\mu}Q_{\mu}. \label{Qnms}
	\end{equation}
With our assumption of nonvanishing curvature $R=0$, we have $-Q \simeq \mathring{R}$, which is thus equivalent to the Einstein-Hilbert Lagrangian.

Since (\ref{Qnms}) is equivalent to GR in the symmetric teleparallel gravity, i.e., in the framework of nonmetricity theory, it is natural to consider modifications of GR by extending (\ref{Qnms}) with more general monomials built of the nonmetricity tensor with couplings to a scalar field. 
At the quadratic order in the nonmetricity tensor, besides the 4 monomials in the nonmetricity scalar (\ref{Qnms}), there is another one $q_{\mu}q^{\mu}$. Generally one may consider the linear combination of these five monomials \cite{Jimenez:2019woj}.
In this work, however, since we concentrate on the modification of GR (i.e., the equivalent nonmetricity scalar $Q$), we simply choose the nonmetricity scalar as the parity preserving term that is quadratic order in the nonmetricity tensor.

In the case of parity violation, the monomials quadratic in the nonmetricity tensor take the schematical form $\epsilon QQ\phi\cdots \phi$, where $\epsilon$ stands for the Levi-Civita tensor $\epsilon_{\mu\nu\rho\sigma} = \sqrt{-g} \varepsilon_{\mu\nu\rho\sigma}$ with $\varepsilon_{0123} = 1$, $Q$ stands for the nonmetricity tensor and $\phi$ stands for the first order derivative of the scalar field $\phi_{\mu} \equiv \nabla_{\mu}\phi$.
We find 6 independent monomials:
\begin{eqnarray}
	\mathcal{F}_{1} & \coloneqq & \epsilon_{\mu\nu\rho\sigma}Q_{\phantom{\mu\nu}\lambda}^{\mu\nu}Q^{\rho\sigma\lambda},\label{calF1}\\
	\mathcal{F}_{2} & \coloneqq & \epsilon_{\mu\nu\rho\sigma}Q_{\phantom{\mu\nu}\alpha}^{\mu\nu}Q_{\phantom{\rho\sigma}\beta}^{\rho\sigma}\phi^{\alpha}\phi^{\beta},\\
	\mathcal{F}_{3} & \coloneqq & \epsilon_{\mu\nu\rho\beta}Q_{\phantom{\mu\nu}\lambda}^{\mu\nu}Q_{\alpha}^{\phantom{\alpha}\rho\lambda}\phi^{\alpha}\phi^{\beta},\\
	\mathcal{F}_{4} & \coloneqq & \epsilon_{\mu\nu\rho\beta}Q_{\phantom{\mu\nu}\lambda}^{\mu\nu}Q_{\phantom{\lambda\rho}\alpha}^{\lambda\rho}\phi^{\alpha}\phi^{\beta},\\
	\mathcal{F}_{5} & \coloneqq & \epsilon_{\mu\nu\rho\beta}Q_{\phantom{\mu\nu}\alpha}^{\mu\nu}Q_{\phantom{\rho\sigma}\sigma}^{\rho\sigma}\phi^{\alpha}\phi^{\beta},\\
	\mathcal{F}_{6} & \coloneqq & \epsilon_{\mu\nu\lambda\beta}Q_{\phantom{\mu\nu}\alpha}^{\mu\nu}Q_{\phantom{\lambda}\rho\sigma}^{\lambda}\phi^{\alpha}\phi^{\beta}\phi^{\rho}\phi^{\sigma}. \label{calF6}
\end{eqnarray}
Note $\mathcal{F}_{1},\cdots,\mathcal{F}_{5}$ have been considered in \cite{Conroy:2019ibo,Li:2022vtn} (where 7 monomials are listed, of which only 5 are independent, see Appendix \ref{app:indep}).
$\mathcal{F}_{2},\cdots,\mathcal{F}_{5}$ are quadratic in the derivative of the scalar field. 
Here we also include $\mathcal{F}_{6}$, which is fourth order in the derivative of the scalar field.
In summary, $\mathcal{F}_{1},\cdots ,\mathcal{F}_{6}$ are the most general parity violating monomials that are quadratic in the nonmetricity tensor with couplings to the first order derivative of a scalar field.

Next we consider the derivative of the nonmetricity tensor $\nabla_{\sigma}Q_{\rho\mu\nu} = \nabla_{\sigma}\nabla_{\rho}g_{\mu\nu}$. 
In order to prevent our model from the ghost problem, we consider monomials that are linear in the derivative of the nonmetricity tensor.
As being stated before, we choose only the nonmetricity scalar $Q$ as the parity preserving term quadratic in the nonmetricity tensor, therefore we do not consider the parity preserving monomials of the form $Q\nabla Q$.
At the lowest order, we focus on the parity violating monomials of the form $\epsilon Q \nabla Q \phi$, where again $\phi$ stands for the first order derivative of the scalar field.
After taking into account the vanishing of curvature tensor, there are 12 contractions \cite{Conroy:2019ibo}:
	\begin{align}
		\mathcal{G}_{1} & =\epsilon^{\mu\nu\rho\sigma}\phi_{\alpha}Q_{\mu\nu}^{\phantom{\mu\nu}\alpha}\nabla_{\beta}Q_{\rho\sigma}^{\phantom{\rho\sigma}\beta}, & \mathcal{G}_{2} & =\epsilon^{\mu\nu\rho\sigma}\phi^{\alpha}Q_{\mu\nu}^{\phantom{\mu\nu}\beta}\nabla_{\alpha}Q_{\rho\sigma\beta},\nonumber \\
		\mathcal{G}_{3} & =\epsilon^{\mu\nu\rho\sigma}\phi^{\alpha}Q_{\mu\nu}^{\phantom{\mu\nu}\beta}\nabla_{\beta}Q_{\rho\sigma\alpha}, & \mathcal{G}_{4} & =\epsilon^{\mu\nu\rho\sigma}\phi_{\mu}q_{\nu}\nabla_{\beta}Q_{\rho\sigma}^{\phantom{\rho\sigma}\beta},\nonumber \\
		\mathcal{G}_{5} & =\epsilon^{\mu\nu\rho\sigma}\phi_{\mu}Q_{\nu}\nabla_{\beta}Q_{\rho\sigma}^{\phantom{\rho\sigma}\beta}, & \mathcal{G}_{6} & =\epsilon^{\mu\nu\rho\sigma}\phi_{\mu}Q_{\phantom{\alpha\beta}\nu}^{\alpha\beta}\nabla_{\alpha}Q_{\rho\sigma\beta},\nonumber \\
		\mathcal{G}_{7} & =\epsilon^{\mu\nu\rho\sigma}\phi_{\mu}Q_{\phantom{\alpha\beta}\nu}^{\alpha\beta}\nabla_{\beta}Q_{\rho\sigma\alpha}, & \mathcal{G}_{8} & =\epsilon^{\mu\nu\rho\sigma}\phi_{\mu}Q_{\nu}^{\phantom{\nu}\alpha\beta}\nabla_{\alpha}Q_{\rho\sigma\beta},\nonumber \\
		\mathcal{G}_{9} & =\epsilon^{\mu\nu\rho\sigma}\phi_{\mu}Q_{\nu\rho}^{\phantom{\nu\rho}\alpha}\nabla_{\alpha}Q_{\beta\phantom{\beta}\sigma}^{\phantom{\beta}\beta}, & \mathcal{G}_{10} & =\epsilon^{\mu\nu\rho\sigma}\phi_{\mu}Q_{\nu\rho}^{\phantom{\nu\rho}\alpha}\nabla_{\alpha}Q_{\sigma\beta}^{\phantom{\sigma\beta}\beta},\nonumber \\
		\mathcal{G}_{11} & =\epsilon^{\mu\nu\rho\sigma}\phi_{\mu}Q_{\nu\rho}^{\phantom{\nu\rho}\alpha}\nabla_{\beta}Q_{\phantom{\beta}\sigma\alpha}^{\beta}, & \mathcal{G}_{12} & =\epsilon^{\mu\nu\rho\sigma}\phi_{\mu}Q_{\nu\rho}^{\phantom{\nu\rho}\alpha}\nabla_{\beta}Q_{\sigma\alpha}^{\phantom{\sigma\alpha}\beta}. \label{calGn}
	\end{align}
In the above, upper indices of $\nabla Q$ are raised by the metric from ``outside'', e.g., $\nabla_{\beta}Q_{\rho\sigma}^{\phantom{\rho\sigma}\beta}\equiv g^{\alpha\beta}\nabla_{\alpha}Q_{\rho\sigma\beta}$, etc.. 
This is to ensure that $\mathcal{G}_n$'s are quadratic order in $Q_{\rho\mu\nu}$.
On the other hand, if one raises indices from ``inside'', since $\nabla_{\rho}g^{\mu\nu}=-g^{\mu\mu'}g^{\nu\nu'}\nabla_{\rho}g_{\mu'\nu'}\equiv-Q_{\rho}^{\phantom{\rho}\mu\nu}$, one has (e.g.)
	\[
	\nabla_{\alpha}\left(Q_{\rho\sigma}^{\phantom{\rho\sigma}\alpha}\right)\equiv\nabla_{\alpha}\left(g^{\alpha\beta}Q_{\rho\sigma\beta}\right)=\nabla_{\alpha}Q_{\rho\sigma}^{\phantom{\rho\sigma}\alpha}-Q_{\alpha}^{\phantom{\alpha}\alpha\beta}Q_{\rho\sigma\beta},
	\]
which indicates that $QQQ$ terms naturally arise at the same order as $Q\nabla Q$ terms. 
This fact also motivates the inclusion of monomials that are cubic order in the nonmetricity tensor, as we shall construct below.

Another motivation of considering monomials cubic order in the nonmetricity tensor comes from the analogue of the ``k-essence'' model of the scalar field theory.
For the Lagrangian of the scalar field in the form $P(X,\phi)$ with $X=-\frac{1}{2}(\partial\phi)^2$ the canonical kinetic term, it is well-known that the non-canonical kinetic term changes the propagating speed of the scalar perturbation to $c_{s}^2 = P_{,X}/(P_{,X}+2X P_{,XX})$.
The same feature appears for the tensor perturbations (e.g.) \cite{Gao:2011vs}.
In a word, higher order terms in the nonmetricity tensor act as the non-canonical kinetic terms for the metric, and will result in a change of the propagating speed of the gravitational waves.

In this work, we concentrate on the monomials that are cubic order in the nonmetricity tensor, which are coupled to the first order derivative of the scalar field. For simplicity, we consider monomials that are linear in $\phi_{\mu}$.
In order to investigate their effect on the propagation of the gravitational waves, we consider both the parity preserving and parity violating cases.
For the parity preserving case, we find 36 monomials of the form $QQQ\phi$:
	\begin{align}
		\mathcal{C}_{1} & =Q_{\alpha\mu\rho}Q_{\phantom{\mu}\nu\sigma}^{\mu}Q^{\rho\nu\sigma}\phi^{\alpha}, & \mathcal{C}_{2} & =Q_{\mu\rho\nu}Q_{\phantom{\mu}\alpha\sigma}^{\mu}Q^{\sigma\rho\nu}\phi^{\alpha}, & \mathcal{C}_{3} & =Q_{\alpha\mu\rho}Q^{\mu}Q^{\rho}\phi^{\alpha},\nonumber \\
		\mathcal{C}_{4} & =Q_{\mu}Q_{\phantom{\mu}\alpha\rho}^{\mu}Q^{\rho}\phi^{\alpha}, & \mathcal{C}_{5} & =Q_{\mu}Q^{\mu}Q_{\alpha}\phi^{\alpha}, & \mathcal{C}_{6} & =Q_{\mu}Q^{\mu}q_{\alpha}\phi^{\alpha},\nonumber \\
		\mathcal{C}_{7} & =Q_{\rho\mu\nu}Q^{\rho\mu\nu}Q_{\alpha}\phi^{\alpha}, & \mathcal{C}_{8} & =Q_{\rho\mu\nu}Q^{\rho\mu\nu}q_{\alpha}\phi^{\alpha}, & \mathcal{C}_{9} & =Q_{\mu\alpha\rho}Q_{\phantom{\rho}\nu\sigma}^{\rho}Q^{\nu\mu\sigma}\phi^{\alpha},\nonumber \\
		\mathcal{C}_{10} & =Q_{\alpha\mu\rho}Q_{\phantom{\mu}\nu\sigma}^{\mu}Q^{\nu\rho\sigma}\phi^{\alpha}, & \mathcal{C}_{11} & =Q_{\mu\nu\sigma}Q_{\phantom{\mu}\alpha\rho}^{\mu}Q^{\nu\rho\sigma}\phi^{\alpha}, & \mathcal{C}_{12} & =Q_{\rho\mu\nu}Q^{\mu\rho\nu}Q_{\alpha}\phi^{\alpha},\nonumber \\
		\mathcal{C}_{13} & =Q_{\rho\mu\nu}Q^{\mu\rho\nu}q_{\alpha}\phi^{\alpha}, & \mathcal{C}_{14} & =Q_{\alpha\mu\rho}Q_{\phantom{\mu\rho}\nu}^{\mu\rho}Q^{\nu}\phi^{\alpha}, & \mathcal{C}_{15} & =Q_{\mu\rho\nu}Q_{\phantom{\mu\rho}\alpha}^{\mu\rho}Q^{\nu}\phi^{\alpha},\nonumber \\
		\mathcal{C}_{16} & =Q_{\mu\alpha\rho}Q_{\phantom{\rho\mu}\nu}^{\rho\mu}Q^{\nu}\phi^{\alpha}, & \mathcal{C}_{17} & =q_{\mu}Q^{\mu}Q_{\alpha}\phi^{\alpha}, & \mathcal{C}_{18} & =q_{\mu}Q^{\mu}q_{\alpha}\phi^{\alpha},\nonumber \\
		\mathcal{C}_{19} & =Q_{\alpha\mu\rho}Q_{\nu}Q^{\nu\mu\rho}\phi^{\alpha}, & \mathcal{C}_{20} & =Q_{\mu\alpha\rho}Q_{\nu}Q^{\nu\mu\rho}\phi^{\alpha}, & \mathcal{C}_{21} & =Q_{\alpha\mu\rho}Q_{\nu\sigma}^{\phantom{\nu\sigma}\rho}Q^{\nu\mu\sigma}\phi^{\alpha},\nonumber \\
		\mathcal{C}_{22} & =Q_{\mu\alpha\rho}Q_{\nu\sigma}^{\phantom{\nu\sigma}\rho}Q^{\nu\mu\sigma}\phi^{\alpha}, & \mathcal{C}_{23} & =Q_{\alpha\mu\rho}Q_{\nu\sigma}^{\phantom{\nu\sigma}\mu}Q^{\sigma\rho\nu}\phi^{\alpha}, & \mathcal{C}_{24} & =Q_{\mu\alpha\rho}Q_{\nu\sigma}^{\phantom{\nu\sigma}\mu}Q^{\sigma\rho\nu}\phi^{\alpha},\nonumber \\
		\mathcal{C}_{25} & =Q_{\mu\alpha\rho}Q^{\rho}q^{\mu}\phi^{\alpha}, & \mathcal{C}_{26} & =Q_{\alpha\mu\rho}Q^{\mu}q^{\rho}\phi^{\alpha}, & \mathcal{C}_{27} & =Q_{\mu}Q_{\phantom{\nu}\alpha\rho}^{\mu}q^{\rho}\phi^{\alpha},\nonumber \\
		\mathcal{C}_{28} & =Q_{\alpha\mu\rho}q^{\mu}q^{\rho}\phi^{\alpha}, & \mathcal{C}_{29} & =Q_{\mu\alpha\rho}q^{\mu}q^{\rho}\phi^{\alpha}, & \mathcal{C}_{30} & =Q_{\alpha\mu\rho}Q_{\phantom{\mu\rho}\nu}^{\mu\rho}q^{\nu}\phi^{\alpha},\nonumber \\
		\mathcal{C}_{31} & =Q_{\mu\rho\nu}Q_{\phantom{\mu\rho}\alpha}^{\mu\rho}q^{\nu}\phi^{\alpha}, & \mathcal{C}_{32} & =Q_{\mu\alpha\rho}Q_{\phantom{\rho\mu}\nu}^{\rho\mu}q^{\nu}\phi^{\alpha}, & \mathcal{C}_{33} & =q_{\mu}q^{\mu}Q_{\alpha}\phi^{\alpha},\nonumber \\
		\mathcal{C}_{34} & =q_{\mu}q^{\mu}q_{\alpha}\phi^{\alpha}, & \mathcal{C}_{35} & =Q_{\alpha\mu\rho}Q_{\nu}^{\phantom{\zeta}\mu\rho}q^{\nu}\phi^{\alpha}, & \mathcal{C}_{36} & =Q_{\mu\alpha\rho}Q_{\nu}^{\phantom{\zeta}\mu\rho}q^{\nu}\phi^{\alpha}. \label{calCn}
	\end{align}
For the parity violating case, we find 44 monomials of the form $\epsilon QQQ\phi$:
\begin{align}
	\mathcal{D}_{1} & =\epsilon^{\mu\nu\rho\sigma}Q_{\phantom{\alpha\beta}\rho}^{\alpha\beta}Q_{\alpha\nu}^{\phantom{\alpha\nu}\delta}Q_{\delta\beta\sigma}\phi_{\mu}, & \mathcal{D}_{2} & =\epsilon^{\mu\nu\rho\sigma}Q_{\phantom{\alpha\beta}\nu}^{\alpha\beta}Q_{\beta\rho}^{\phantom{\beta\rho}\delta}Q_{\delta\alpha\sigma}\phi_{\mu}, & \mathcal{D}_{3} & =\epsilon^{\mu\nu\rho\sigma}Q_{\nu}^{\phantom{\nu}\alpha\beta}Q_{\alpha\rho}^{\phantom{\alpha\rho}\delta}Q_{\delta\beta\sigma}\phi_{\mu},\nonumber \\
	\mathcal{D}_{4} & =\epsilon^{\mu\nu\rho\sigma}Q_{\mu\nu}^{\phantom{\mu\nu}\alpha}Q_{\rho}^{\phantom{\rho}\beta\delta}Q_{\beta\delta\sigma}\phi_{\alpha}, & \mathcal{D}_{5} & =\epsilon^{\mu\nu\rho\sigma}q_{\nu}Q_{\rho}^{\phantom{\rho}\delta\beta}Q_{\beta\delta\sigma}\phi_{\mu}, & \mathcal{D}_{6} & =\epsilon^{\mu\nu\rho\sigma}Q_{\nu}Q_{\rho}^{\phantom{\rho}\alpha\beta}Q_{\alpha\beta\sigma}\phi_{\mu},\nonumber \\
	\mathcal{D}_{7} & =\epsilon^{\mu\nu\rho\sigma}Q_{\alpha}^{\phantom{\alpha}\beta\delta}Q_{\beta\delta\nu}Q_{\rho\sigma}^{\phantom{\rho\sigma}\alpha}\phi_{\mu}, & \mathcal{D}_{8} & =\epsilon^{\mu\nu\rho\sigma}Q_{\alpha}^{\phantom{\alpha}\beta\delta}Q_{\mu\nu\beta}Q_{\rho\sigma\delta}\phi^{\alpha}, & \mathcal{D}_{9} & =\epsilon^{\mu\nu\rho\sigma}Q_{\beta\delta\alpha}Q_{\mu\nu}^{\phantom{\mu\nu}\beta}Q_{\rho\sigma}^{\phantom{\rho\sigma}\delta}\phi^{\alpha},\nonumber \\
	\mathcal{D}_{10} & =\epsilon^{\mu\nu\rho\sigma}Q_{\alpha\mu}^{\phantom{\alpha\mu}\beta}Q_{\beta\nu}^{\phantom{\beta\nu}\delta}Q_{\rho\sigma\delta}\phi^{\alpha}, & \mathcal{D}_{11} & =\epsilon^{\mu\nu\rho\sigma}Q_{\phantom{\beta\delta}\nu}^{\beta\delta}Q_{\beta\mu}^{\phantom{\beta\mu}\alpha}Q_{\rho\sigma\delta}\phi_{\alpha}, & \mathcal{D}_{12} & =\epsilon^{\mu\nu\rho\sigma}Q_{\alpha\nu}^{\phantom{\alpha\nu}\delta}Q^{\alpha}Q_{\rho\sigma\delta}\phi_{\mu},\nonumber \\
	\mathcal{D}_{13} & =\epsilon^{\mu\nu\rho\sigma}Q_{\alpha\beta\nu}Q^{\alpha\beta\delta}Q_{\rho\delta\sigma}\phi_{\mu}, & \mathcal{D}_{14} & =\epsilon^{\mu\nu\rho\sigma}Q_{\alpha}^{\phantom{\alpha}\delta\beta}Q_{\beta\nu}^{\phantom{\beta\nu}\alpha}Q_{\rho\sigma\delta}\phi_{\mu}, & \mathcal{D}_{15} & =\epsilon^{\mu\nu\rho\sigma}Q_{\mu}^{\phantom{\mu}\beta\alpha}Q_{\beta\nu}^{\phantom{\beta\nu}\delta}Q_{\rho\sigma\delta}\phi_{\alpha},\nonumber \\
	\mathcal{D}_{16} & =\epsilon^{\mu\nu\rho\sigma}q^{\beta}Q_{\beta\nu}^{\phantom{\beta\nu}\delta}Q_{\rho\sigma\delta}\phi_{\mu}, & \mathcal{D}_{17} & =\epsilon^{\mu\nu\rho\sigma}Q_{\alpha}Q_{\mu\nu}^{\phantom{\mu\nu}\delta}Q_{\rho\sigma\delta}\phi^{\alpha}, & \mathcal{D}_{18} & =\epsilon^{\mu\nu\rho\sigma}q_{\alpha}Q_{\mu\nu}^{\phantom{\mu\nu}\delta}Q_{\rho\sigma\delta}\phi^{\alpha},\nonumber \\
	\mathcal{D}_{19} & =\epsilon^{\mu\nu\rho\sigma}Q_{\alpha\mu}^{\phantom{\alpha\mu}\beta}Q_{\delta\beta\nu}Q_{\rho\sigma}^{\phantom{\rho\sigma}\delta}\phi^{\alpha}, & \mathcal{D}_{20} & =\epsilon^{\mu\nu\rho\sigma}Q_{\beta\alpha\mu}Q_{\delta\nu}^{\phantom{\delta\nu}\beta}Q_{\rho\sigma}^{\phantom{\rho\sigma}\delta}\phi^{\alpha}, & \mathcal{D}_{21} & =\epsilon^{\mu\nu\rho\sigma}Q_{\beta}Q_{\delta\nu}^{\phantom{\delta\nu}\beta}Q_{\rho\sigma}^{\phantom{\rho\sigma}\delta}\phi_{\mu},\nonumber \\
	\mathcal{D}_{22} & =\epsilon^{\mu\nu\rho\sigma}Q_{\mu\alpha\beta}Q_{\delta\nu}^{\phantom{\delta\nu}\beta}Q_{\rho\sigma}^{\phantom{\rho\sigma}\delta}\phi^{\alpha}, & \mathcal{D}_{23} & =\epsilon^{\mu\nu\rho\sigma}q_{\beta}Q_{\delta\nu}^{\phantom{\delta\nu}\beta}Q_{\rho\sigma}^{\phantom{\rho\sigma}\delta}\phi_{\mu}, & \mathcal{D}_{24} & =\epsilon^{\mu\nu\rho\sigma}Q_{\mu\nu\alpha}Q_{\delta}Q_{\rho\sigma}^{\phantom{\rho\sigma}\delta}\phi^{\alpha},\nonumber \\
	\mathcal{D}_{25} & =\epsilon^{\mu\nu\rho\sigma}q_{\nu}Q_{\delta}Q_{\rho\sigma}^{\phantom{\rho\sigma}\delta}\phi_{\mu}, & \mathcal{D}_{26} & =\epsilon^{\mu\nu\rho\sigma}Q_{\nu\delta\alpha}Q_{\phantom{\alpha\delta}\beta}^{\alpha\delta}Q_{\rho\sigma}^{\phantom{\rho\sigma}\beta}\phi_{\mu}, & \mathcal{D}_{27} & =\epsilon^{\mu\nu\rho\sigma}Q_{\alpha\mu\beta}Q_{\nu\delta}^{\phantom{\nu\delta}\beta}Q_{\rho\sigma}^{\phantom{\rho\sigma}\delta}\phi^{\alpha},\nonumber \\
	\mathcal{D}_{28} & =\epsilon^{\mu\nu\rho\sigma}Q_{\beta\alpha\mu}Q_{\nu\delta}^{\phantom{\nu\delta}\beta}Q_{\rho\sigma}^{\phantom{\rho\sigma}\delta}\phi^{\alpha}, & \mathcal{D}_{29} & =\epsilon^{\mu\nu\rho\sigma}Q_{\alpha}Q_{\nu\delta}^{\phantom{\nu\delta}\alpha}Q_{\rho\sigma}^{\phantom{\rho\sigma}\delta}\phi_{\mu}, & \mathcal{D}_{30} & =\epsilon^{\mu\nu\rho\sigma}Q_{\mu\beta}^{\phantom{\mu\beta}\alpha}Q_{\nu\delta}^{\phantom{\nu\delta}\beta}Q_{\rho\sigma}^{\phantom{\rho\sigma}\delta}\phi_{\alpha},\nonumber \\
	\mathcal{D}_{31} & =\epsilon^{\mu\nu\rho\sigma}q_{\beta}Q_{\nu\delta}^{\phantom{\nu\delta}\beta}Q_{\rho\sigma}^{\phantom{\rho\sigma}\delta}\phi_{\mu}, & \mathcal{D}_{32} & =\epsilon^{\mu\nu\rho\sigma}Q_{\mu\nu}^{\phantom{\mu\nu}\alpha}q^{\delta}Q_{\rho\sigma\delta}\phi_{\alpha}, & \mathcal{D}_{33} & =\epsilon^{\mu\nu\rho\sigma}Q_{\alpha\mu}^{\phantom{\alpha\mu}\beta}Q_{\nu\beta\rho}q_{\sigma}\phi^{\alpha},\nonumber \\
	\mathcal{D}_{34} & =\epsilon^{\mu\nu\rho\sigma}Q_{\beta\alpha\mu}Q_{\nu\rho}^{\phantom{\nu\rho}\beta}q_{\sigma}\phi^{\alpha}, & \mathcal{D}_{35} & =\epsilon^{\mu\nu\rho\sigma}Q_{\mu\alpha\beta}Q_{\nu\rho}^{\phantom{\nu\rho}\beta}q_{\sigma}\phi^{\alpha}, & \mathcal{D}_{36} & =\epsilon^{\mu\nu\rho\sigma}q_{\beta}Q_{\nu\rho}^{\phantom{\nu\rho}\beta}q_{\sigma}\phi_{\mu},\nonumber \\
	\mathcal{D}_{37} & =\epsilon^{\mu\nu\rho\sigma}Q_{\alpha\beta\mu}Q_{\nu\rho}^{\phantom{\nu\rho}\beta}Q_{\sigma}\phi^{\alpha}, & \mathcal{D}_{38} & =\epsilon^{\mu\nu\rho\sigma}Q_{\beta\alpha\mu}Q_{\nu\rho}^{\phantom{\nu\rho}\beta}Q_{\sigma}\phi^{\alpha}, & \mathcal{D}_{39} & =\epsilon^{\mu\nu\rho\sigma}Q_{\alpha}Q_{\nu\rho}^{\phantom{\nu\rho}\alpha}Q_{\sigma}\phi_{\mu},\nonumber \\
	\mathcal{D}_{40} & =\epsilon^{\mu\nu\rho\sigma}Q_{\mu\alpha\beta}Q_{\nu\rho}^{\phantom{\nu\rho}\beta}Q_{\sigma}\phi^{\alpha}, & \mathcal{D}_{41} & =\epsilon^{\mu\nu\rho\sigma}q_{\beta}Q_{\nu\rho}^{\phantom{\nu\rho}\beta}Q_{\sigma}\phi_{\mu}, & \mathcal{D}_{42} & =\epsilon^{\mu\nu\rho\sigma}Q_{\mu\nu\alpha}q_{\rho}Q_{\sigma}\phi^{\alpha},\nonumber \\
	\mathcal{D}_{43} & =\epsilon^{\mu\nu\rho\sigma}Q_{\nu\alpha\beta}Q_{\phantom{\alpha\delta}\rho}^{\alpha\delta}Q_{\sigma\delta}^{\phantom{\sigma\delta}\beta}\phi_{\mu}, & \mathcal{D}_{44} & =\epsilon^{\mu\nu\rho\sigma}Q_{\nu\rho}^{\phantom{\nu\rho}\alpha}Q_{\alpha}^{\phantom{\alpha}\delta\beta}Q_{\sigma\beta\delta}\phi_{\mu}. \label{calDn}
\end{align}
We emphasize that in constructing the above monomials, we only made use of the (anti-)symmetries of the $\epsilon$-tensor and the nonmetricity tensor. In particular, we have not examined if additional relations among these monomials exist by using the identity (\ref{antisymm_id}).

Combining together all the monomials in the above, the full action we shall consider is given by
\begin{equation}
	S=\int\mathrm{d}^{4}x\sqrt{-g}\left(-\frac{Q}{2}+\sum_{n=1}^{6}f_{n}\mathcal{F}_{n}+\sum_{n=1}^{12}g_{n}\mathcal{G}_{n}+\sum_{n=1}^{36}c_{n}\mathcal{C}_{n}+\sum_{n=1}^{44}d_{n}\mathcal{D}_{n}-\frac{1}{2}\partial_{\mu}\phi\partial^{\mu}\phi-V\left(\phi\right)\right), \label{action}
\end{equation}
where $\mathcal{F}_{n}$, $\mathcal{G}_{n}$, $\mathcal{C}_{n}$ and $\mathcal{D}_{n}$ are the scalar-nonmetricity monomials listed in the above, and the coefficients $f_{n}$, $g_{n}$, $c_{n}$ and $d_{n}$ are functions of the scalar field $\phi$ only.
In (\ref{action}) we have also introduced a canonical kinetic terms for the scalar field, while neglected other matter content in the universe.

\section{Polarization of the gravitational waves} \label{sec:gws}

In this section, we investigate the propagation of the gravitational waves in the model (\ref{action}) on a cosmological background.
Since only the tensor perturbations are involved, the perturbed metric can be parametrized by
	\begin{equation}
		\mathrm{d}s^{2}=-\mathrm{d}t^{2}+a^{2}\left(t\right)\mathfrak{g}_{ij}\mathrm{d}x^{i}\mathrm{d}x^{j},
	\end{equation}
where $\mathfrak{g}_{ij}\equiv\delta_{ik}\left(e^{\bm{\gamma}}\right)_{\phantom{k}j}^{k}$ with 
	\begin{equation}
		\left(e^{\bm{\gamma}}\right)_{\phantom{i}j}^{i}=\delta_{\phantom{i}j}^{i}+\gamma_{\phantom{i}j}^{i}+\frac{1}{2}\gamma_{\phantom{i}k}^{i}\gamma_{\phantom{k}j}^{k}+\cdots.
	\end{equation}
Here $\gamma_{\phantom{i}j}^{i}$ are the transverse and traceless tensor perturbations satisfying $\gamma_{\phantom{i}i}^{i}=0$ and $\partial_{i}\gamma_{\phantom{i}j}^{i}=0$.
In the following, spatial indices are raised and lowered by $\delta^{ij}$ and $\delta_{ij}$.

When the nonmetricity tensor is present, the associated affine connection is independent of the spacetime metric. In our case with vanishing curvature and torsion tensors, the affine connection is given by (\ref{Gamma}). Therefore the four scalar field $\xi^{a}$ are treated as independent variables, which must be determined by their own equations of motion.
Nevertheless, in the cosmological background it is possible to have the solution $\xi^{a} \propto \delta^{a}_{\mu}x^{\mu}$ such that the affine connection is identically vanishing.
Such a choice of a vanishing affine connection is dubbed the ``coincident gauge'' in the literature \cite{BeltranJimenez:2017tkd,Zhao:2021zab}.
In the following we shall work with the coincident gauge, in which the nonmetricity tensor reduces to be $Q_{\rho\mu\nu} = \partial_{\rho}g_{\mu\nu}$.
In practice, since in this paper we consider only the linear tensor perturbations, a general affine connection the form (\ref{Gamma}) actually only affect the scalar and vector perturbation, and will not contribute to the tensor perturbations.

Our next task is to derive the quadratic action for the tensor perturbations in the cosmological background.
Not all the monomials in (\ref{action}) will contribute to the tensor perturbations. 
For the $\mathcal{F}_{n}$'s, only $\mathcal{F}_{1}$ and $\mathcal{F}_{3}$ contribute to the tensor perturbations:
\begin{eqnarray}
	\mathcal{F}_{1} & \supset & -2a^{2}\varepsilon^{ijk}\dot{\gamma}_{\phantom{l}i}^{l}\partial_{j}\gamma_{lk},\\
	\mathcal{F}_{3} & \supset & -a^{2}\dot{\phi}^{2}\varepsilon^{ijk}\dot{\gamma}_{\phantom{l}i}^{l}\partial_{j}\gamma_{lk},
\end{eqnarray}
where $\varepsilon^{ijk}$ is the 3-dimension Levi-Civita symbol with $\varepsilon_{123}= \varepsilon^{123} = 1$.
In the above no integration by parts has been made.
Similarly, the contributions from the $\mathcal{G}_{n}$'s are
\begin{eqnarray}
	\mathcal{G}_{2} & \supset & -6a\dot{a}\dot{\phi}\varepsilon^{ijk}\partial_{j}\gamma_{\phantom{l}i}^{l}\dot{\gamma}_{lk}-a^{2}\dot{\phi}\varepsilon^{ijk}\dot{\gamma}_{\phantom{l}i}^{l}\partial_{k}\dot{\gamma}_{lj}-a^{2}\dot{\phi}\varepsilon^{ijk}\partial_{j}\gamma_{\phantom{l}i}^{l}\ddot{\gamma}_{lk},\\
	\mathcal{G}_{6} & \supset & \dot{\phi}\varepsilon^{ijk}\partial_{k}\partial^{l}\gamma_{mj}\partial_{l}\gamma_{\phantom{m}i}^{m}-2a\dot{a}\dot{\phi}\varepsilon^{ijk}\partial_{j}\gamma_{\phantom{l}i}^{l}\dot{\gamma}_{lk}-a^{2}\dot{\phi}\varepsilon^{ijk}\dot{\gamma}_{\phantom{l}i}^{l}\partial_{k}\dot{\gamma}_{lj}, \label{calG6}\\
	\mathcal{G}_{11} & \supset & \dot{\phi}\varepsilon^{ijk}\partial_{j}\gamma_{\phantom{l}i}^{l}\partial^{2}\gamma_{lk}-4a\dot{a}\dot{\phi}\varepsilon^{ijk}\partial_{j}\gamma_{\phantom{l}i}^{l}\dot{\gamma}_{lk}-a^{2}\dot{\phi}\varepsilon^{ijk}\partial_{j}\gamma_{\phantom{l}i}^{l}\ddot{\gamma}_{lk}. \label{calG11}
\end{eqnarray}
For monomials that are cubic in the nonmetricity tensor, we find, for the parity preserving contributions:
\begin{eqnarray}
	\mathcal{C}_{1} & \supset & -2\dot{a}\dot{\phi}\partial_{i}\gamma_{jk}\partial^{i}\gamma^{jk},\\
	\mathcal{C}_{7} & \supset & -6\dot{a}\dot{\phi}\partial_{i}\gamma_{jk}\partial^{i}\gamma^{jk}+6a^{2}\dot{a}\dot{\phi}\dot{\gamma}_{ij}\dot{\gamma}^{ij},\\
	\mathcal{C}_{19} & \supset & 6a^{2}\dot{a}\dot{\phi}\dot{\gamma}_{ij}\dot{\gamma}^{ij},\\
	\mathcal{C}_{21} & \supset & -2\dot{a}\dot{\phi}\partial_{i}\gamma_{jk}\partial^{i}\gamma^{jk}+6a^{2}\dot{a}\dot{\phi}\dot{\gamma}_{ij}\dot{\gamma}^{ij},
\end{eqnarray}
and for the parity violating contributions:
\begin{eqnarray}
	\mathcal{D}_{1} & \supset & 2a\dot{a}\dot{\phi}\varepsilon^{ijk}\partial_{j}\gamma_{li}\dot{\gamma}_{\phantom{l}k}^{l},\\
	\mathcal{D}_{8} & \supset & -8a\dot{a}\dot{\phi}\varepsilon^{ijk}\partial_{j}\gamma_{li}\dot{\gamma}_{\phantom{l}k}^{l},\\
	\mathcal{D}_{10} & \supset & -2a\dot{a}\dot{\phi}\varepsilon^{ijk}\partial_{j}\gamma_{li}\dot{\gamma}_{\phantom{l}k}^{l},\\
	\mathcal{D}_{12} & \supset & -6a\dot{a}\dot{\phi}\varepsilon^{ijk}\partial_{j}\gamma_{li}\dot{\gamma}_{\phantom{l}k}^{l},\\
	\mathcal{D}_{13} & \supset & -4a\dot{a}\dot{\phi}\varepsilon^{ijk}\partial_{j}\gamma_{li}\dot{\gamma}_{\phantom{l}k}^{l},\\
	\mathcal{D}_{17} & \supset & -12a\dot{a}\dot{\phi}\varepsilon^{ijk}\partial_{j}\gamma_{li}\dot{\gamma}_{\phantom{l}k}^{l},\\
	\mathcal{D}_{19} & \supset & 2a\dot{a}\dot{\phi}\varepsilon^{ijk}\partial_{j}\gamma_{li}\dot{\gamma}_{\phantom{l}k}^{l},\\
	\mathcal{D}_{27} & \supset & 4a\dot{a}\dot{\phi}\varepsilon^{ijk}\partial_{j}\gamma_{li}\dot{\gamma}_{\phantom{l}k}^{l},\\
	\mathcal{D}_{37} & \supset & 6a\dot{a}\dot{\phi}\varepsilon^{ijk}\partial_{j}\gamma_{li}\dot{\gamma}_{\phantom{l}k}^{l}.
\end{eqnarray}

Putting all the contributions together and after performing integrations by parts, the full quadratic action for the tensor perturbations is found to be
	\begin{eqnarray}
		S_{\text{T}}^{\left(2\right)} & = & \int\mathrm{d}t\frac{\mathrm{d}^{3}k}{\left(2\pi\right)^{3}}\frac{a^{3}}{2}\bigg[\mathcal{G}_{0}\dot{\gamma}^{ij}\left(t,\bm{k}\right)\dot{\gamma}_{ij}\left(t,-\bm{k}\right)+\mathcal{G}_{1}\varepsilon^{ijk}\left(\frac{-ik_{j}}{a}\right)\dot{\gamma}_{\phantom{l}i}^{l}\left(t,\bm{k}\right)\dot{\gamma}_{lk}\left(t,-\bm{k}\right)\nonumber \\
		&  & \qquad\qquad -\mathcal{W}_{0}\frac{k^{2}}{a^{2}}\gamma_{ij}\left(t,\bm{k}\right)\gamma^{ij}\left(t,-\bm{k}\right)-\mathcal{W}_{-1}\left(\frac{-ik_{j}}{a}\right)\varepsilon^{ijk}\gamma_{\phantom{l}i}^{l}\left(t,\bm{k}\right)\gamma_{lk}\left(t,-\bm{k}\right)\nonumber \\
		&  & \qquad\qquad -\mathcal{W}_{1}\varepsilon^{ijk}\gamma_{\phantom{l}i}^{l}\left(t,\bm{k}\right)\frac{k^{2}}{a^{2}}\left(\frac{-ik_{j}}{a}\right)\gamma_{lk}\left(t,-\bm{k}\right)\bigg], \label{ST2}
	\end{eqnarray}
where we have moved to the Fourier space and the coefficients $\mathcal{G}_{0}$, $\mathcal{W}_{0}$ etc. are given by
	\begin{eqnarray}
		\mathcal{G}_{0} & = & \frac{1}{4}+12\left(c_{7}+c_{19}+c_{21}\right)H\dot{\phi}, \label{calG0}\\
		\mathcal{W}_{0} & = & \frac{1}{4}+4\left(c_{1}+3c_{7}+c_{21}\right)H\dot{\phi}, \label{calW0}\\
		\mathcal{W}_{-1} & = & -2\dot{f}_{1}-4f_{1}H-\dot{f}_{3}\dot{\phi}^{2}-2f_{3}H\dot{\phi}^{2}-2f_{3}\dot{\phi}\ddot{\phi}\nonumber \\
		&  & +\frac{2}{a^{2}}\partial_{t}\left[\dot{\phi}a^{2}H\left(2g_{2}+g_{6}+g_{11}\right)-\frac{1}{2}\dot{\phi}a^{2}\left(\dot{g}_{2}+\dot{g}_{11}\right)-\frac{1}{2}a^{2}\ddot{\phi}_{0}\left(g_{2}+g_{11}\right)\right]\nonumber \\
		&  & -\frac{2}{a^{2}}\partial_{t}\left[\left(d_{1}-4d_{8}-d_{10}-3d_{12}-2d_{13}-6d_{17}+d_{19}+2d_{27}+3d_{37}\right)a^{2}H\dot{\phi}\right],\\
		\mathcal{G}_{1}=\mathcal{W}_{1} & = & -2\dot{\phi}\left(g_{6}-g_{11}\right). \label{calG1calW1}
	\end{eqnarray}
In (\ref{ST2}), terms proportional to $\mathcal{G}_{0}$ and $\mathcal{W}_{0}$ take the same form as those in GR, which are parity preserving. The standard results in GR, i.e., with all $\mathcal{F}_{n} = \mathcal{G}_{n} = \mathcal{C}_{n} = \mathcal{D}_{n} = 0$, correspond to $\mathcal{G}_{0} = \mathcal{W}_{0} = \frac{1}{4}$.
Terms proportional to $\mathcal{G}_{1}$, $\mathcal{W}_{-1}$ and $\mathcal{W}_{1}$ involve $\varepsilon^{ijk}$, which signal the parity violating effects in the tensor perturbations.
The quadratic action (\ref{ST2}) explicitly shows that there is neither extra polarization modes of the GWs nor mixing between the two polarization modes in the parity violating scalar-nonmetricity theory (\ref{action}).

The quadratic action (\ref{ST2}) takes the general form in \cite{Gao:2019liu}, which is based on the framework of spatially covariant gravity.
While comparing with the case of spatially covariant gravity with parity violation, here there arises a contribution proportional to $\mathcal{W}_{-1}$, which is first order in the spatial derivatives (i.e., linear in $\bm{k}$ in the Fourier space).
Such contributions have been reported in \cite{Conroy:2019ibo,Li:2021mdp,Li:2022vtn}, as well as in the NY modified gravity with torsion \cite{Li:2020xjt,Li:2021wij,Wu:2021ndf,Cai:2021uup}.
Such contributions come from the terms with ``mixed'' temporal and spatial derivatives, which can be reduced by integration by parts using
	\begin{equation}
		f\left(t\right)\varepsilon^{ijk}\partial_{j}\gamma_{\phantom{l}i}^{l}\dot{\gamma}_{lk}\simeq-\frac{1}{2}\dot{f}\left(t\right)\varepsilon^{ijk}\partial_{j}\gamma_{\phantom{l}i}^{l}\gamma_{lk},
	\end{equation}
and yield the terms with a single spatial derivative.
It is interesting that such kind of terms does not arise within the framework of spatially covariant gravity considered in \cite{Gao:2019liu}.
While such contributions are generally present in the parity violating scalar-nonmetricity theory. This can be seen form $\mathcal{W}_{-1}$ that it receives contributions from all types of $\mathcal{F}_{n}$, $\mathcal{G}_{n}$ and $\mathcal{D}_{n}$ monomials.
As a result, such kind of terms as well as their imprint on the gravitational waves can be regarded as a characteristic feature of parity violating scalar-nonmetricity theories.

The tensor perturbations $\gamma_{ij}$ can be decomposed into the polarization modes
\begin{equation}
	\gamma_{ij}\left(t,\bm{k}\right)=\sum_{s=\pm2}e_{ij}^{\left(s\right)}(\hat{\bm{k}})\gamma^{\left(s\right)}\left(t,\bm{k}\right),
\end{equation}
where $\hat{\bm{k}} = \frac{\bm{k}}{|\bm{k}|}$, $e_{ij}^{\left(s\right)}(\hat{\bm{k}})$ are the circular polarization tensors with $s=\pm 2$ the helicity states. The polarization tensors satisfy the transverse and traceless conditions
	\begin{equation}
		\delta^{ij}e_{ij}^{\left(s\right)}(\hat{\bm{k}})=0,\quad k^{i}e_{ij}^{\left(s\right)}(\hat{\bm{k}})=0.
	\end{equation}
In order to fully determine the polarization tensors, we choose the phase of $e_{ij}^{\left(s\right)}(\hat{\bm{k}})$ such that \cite{Gao:2011vs} 
	\begin{equation}
	e_{ij}^{\left(s\right)*}(\hat{\bm{k}})=e_{ij}^{-s}(\hat{\bm{k}})=e_{ij}^{\left(s\right)}(-\hat{\bm{k}}),
	\end{equation}
where an asterisk stands for the complex conjugate. The polarization tensors are normalized to be
\begin{equation}
	e_{ij}^{\left(s\right)}(\hat{\bm{k}})e^{\left(-s'\right)ij}(\hat{\bm{k}})=\delta^{ss'}.
\end{equation}
With these settings, there is a useful relation \cite{Alexander:2004wk,Satoh:2007gn,Bartolo:2017szm}
\begin{equation}
	i\hat{k}^{l}\varepsilon_{lij}e_{m}^{\left(s\right)i}(\hat{\bm{k}})e^{\left(s'\right)jm}(-\hat{\bm{k}})=\frac{s}{2}\delta^{ss'}.
\end{equation}

After some manipulations, the full quadratic action for the circular polarization modes is
	\begin{eqnarray}
		S_{\mathrm{T}}^{\left(2\right)} & = & \int\mathrm{d}\tau\frac{\mathrm{d}^{3}k}{\left(2\pi\right)^{3}}\sum_{s=\pm2}\frac{a^{2}}{2}\bigg[\mathcal{G}^{\left(s\right)}\left(\tau,k\right)\gamma'^{\left(s\right)}\left(\tau,\bm{k}\right)\gamma'^{\left(s\right)}\left(\tau,-\bm{k}\right)\nonumber \\
		&  & \qquad-k^{2}\mathcal{W}^{\left(s\right)}\left(\tau,k\right)\gamma^{\left(s\right)}\left(\tau,\bm{k}\right)\gamma^{\left(s\right)}\left(\tau,-\bm{k}\right)\bigg], \label{ST2p}
	\end{eqnarray}
where we have used the conformal time $\tau$ with $\mathrm{d}t = a \mathrm{d}\tau$, and a prime denotes derivative with respect to $\tau$.
Following \cite{Gao:2019liu}, in (\ref{ST2p}) we have defined
	\begin{eqnarray}
		\mathcal{G}^{\left(s\right)}\left(\tau,k\right) & = & \sum_{n=0}^{1}\mathcal{G}_{n}\left(\tau\right)\left(\frac{sk}{2a}\right)^{n}=\mathcal{G}_{0}\left(\tau\right)+\mathcal{G}_{1}\left(\tau\right)\frac{sk}{2a},\\
		\mathcal{W}^{\left(s\right)}\left(\tau,k\right) & = & \sum_{n=-1}^{1}\mathcal{W}_{n}\left(\tau\right)\left(\frac{sk}{2a}\right)^{n}=\mathcal{W}_{-1}\left(\tau\right)\frac{sa}{2k}+\mathcal{W}_{0}\left(\tau\right)+\mathcal{W}_{1}\left(\tau\right)\frac{sk^{3}}{2a^{3}}.
	\end{eqnarray}
We emphasize that the coefficients $\mathcal{G}_{0}(\tau)$ and  $\mathcal{G}_{1}(\tau)$ should not be confused with the monomials $\mathcal{G}_{n}$'s in (\ref{calGn}).

The equations of motion for the circularly polarized modes are
\begin{equation}
	\gamma''{}^{\left(s\right)}\left(\tau,\bm{k}\right)+\mathcal{H}\left(2+\nu^{\left(s\right)}\right)\gamma'{}^{\left(s\right)}\left(\tau,\bm{k}\right)+\left(c_{\text{T}}^{\left(s\right)}\right)^{2}k^{2}\gamma^{\left(s\right)}\left(\tau,\bm{k}\right)=0, \quad s=\pm 2, \label{eom}
\end{equation}
where $\mathcal{H} = \frac{a'}{a}$ is the comoving Hubble parameter,
	\begin{equation}
			\nu^{\left(s\right)}\left(\tau,k\right)  =\frac{1}{\mathcal{H}}\frac{\partial_{\tau}\mathcal{G}^{\left(s\right)}\left(\tau,k\right)}{\mathcal{G}{}^{\left(s\right)}\left(\tau,k\right)}, \label{nu}
	\end{equation}
and
	\begin{equation}
		\left(c_{\mathrm{T}}^{\left(s\right)}\right)^{2}=\frac{\mathcal{W}_{-1}\frac{2a}{sk}+\mathcal{W}_{0}+\mathcal{W}_{1}\frac{sk}{2a}}{\mathcal{G}_{0}+\mathcal{G}_{1}\frac{sk}{2a}}. \label{cT2}
	\end{equation}
The parameter $\nu^{(s)}$ defined in (\ref{nu}) is running rate of the effective Planck mass  \cite{Lagos:2019kds}, $c_{\mathrm{T}}^{(s)}$ is regarded as the propagating speed of the GWs \cite{Saltas:2014dha,Sawicki:2016klv,Nishizawa:2017nef}.
Generally, due to the presence of parity violating terms in the action (\ref{action}), $\nu^{(+2)} \neq \nu^{(-2)}$ and thus the left/right-hand polarizations of the GWs acquire different dampings, which is the effect of ``amplitude birefringence'' \cite{Alexander:2004wk,Yunes:2010yf,Yagi:2012ya,Alexander:2017jmt,Yagi:2017zhb}.
On the other hand, $c_{\mathrm{T}}^{(+2)} \neq c_{\mathrm{T}}^{(-2)}$ implies that the left/right-hand polarizations of the GWs propagate with different velocities, which is the effect of ``velocity birefringence'' \cite{Takahashi:2009wc,Wang:2012fi,Zhu:2013fja,Nishizawa:2018srh}.

From (\ref{calG0}) and (\ref{calG1calW1}), $\mathcal{G}_{0}(\tau)$ and $\mathcal{G}_{1}(\tau)$ receive contributions from $\mathcal{G}_{n}$ and $\mathcal{C}_{n}$ terms in the action (\ref{action}), while not from $\mathcal{F}_{n}$ terms.
In other words, the parameter $\nu^{(s)}$ is vanishing if only the $\mathcal{F}_{n}$ terms are present (besides the nonmetricity scalar $Q$).
This is also consistent with the result in \cite{Conroy:2019ibo}.
From (\ref{calG0}) and (\ref{calW0}), it is interesting to note that even without the parity violating terms (i.e., by setting $\mathcal{G}_{n} = \mathcal{D}_{n}=0$), $\mathcal{G}_{0}(\tau)$ and $\mathcal{W}_{0}(\tau)$ will receive modifications from $\mathcal{C}_{n}$'s, i.e., monomials of the form $QQQ$.
This justifies our motivation of introducing monomials of the form $QQQ$, which act as the non-canonical (non-quadratic) kinetic terms for the metric as the analogue of ``k-essence'', and result in the change of the propagating speed of the GWs.

Due to the modification of $c_{\mathrm{T}}^{(s)}$, generally the polarization modes of the GWs may suffer from the gradient instability if $\left(c_{\mathrm{T}}^{\left(s\right)}\right)^{2} <0 $.
In particular, for long wavelength modes, the term $\mathcal{W}_{-1}\frac{2a}{sk}$ in $\left(c_{\mathrm{T}}^{\left(s\right)}\right)^{2}$ dominates. 
Without loss of generality, we assume $\mathcal{W}_{-1}>0$ and thus the left-handed polarization mode (with $s=-2$) experiences a gradient instability since $\left(c_{\mathrm{T}}^{\left(-2\right)}\right)^{2} \approx -\mathcal{W}_{-1} \frac{a}{k}<0$, which yields an exponential growth of the perturbation mode.
The situation becomes even worse as the wavelength goes larger, i.e., on the large scales.
As we have argued before, the contributions to $\mathcal{W}_{-1}$ are generally present in the parity violating scalar-nonmetricity theory, therefore such a gradient instability is a characteristic feature of the parity violating scalar-nonmetricity theory.

\section{Conclusion} \label{sec:con}

Parity violating features in the gravity theory, in particular in the gravitational waves, have attracted much attention recently.
In this work, we consider a class of scalar-nonmetricity gravity theory with parity violation, and investigate the propagation of the gravitational waves in such kind of theory.
The Lagrangian of the scalar-nonmetricity theory is a polynomial built of the nonmetricity tensor $Q_{\rho\mu\nu}$ coupled with the first order derivative of a scalar field $\phi$.
The parity violating monomials in the form $\sim\epsilon QQ$ and $\sim\epsilon Q\nabla Q$ have been considered in \cite{Conroy:2019ibo}.
In this work, instead of considering higher derivative terms, we consider ``k-essence''-like generalization, i.e., higher order in the nonmetricity tensor $Q_{\rho\mu\nu}$.
Such kind of monomials, being first order in the spacetime metric, are manifestly ghostfree.
For completeness, we build monomials in both the parity preserving and parity violating cases, which are of the form $\sim QQQ\phi$ given in $\mathcal{C}_{n}$'s in (\ref{calCn}) and of the form $\sim \epsilon QQQ\phi$ given in $\mathcal{D}_{n}$'s in (\ref{calDn}), respectively.

The full action of our scalar-nonmetricity theory is given in (\ref{action}). We then investigated the tensor perturbations of this model in a cosmological background.
By deriving the quadratic action for the tensor perturbations (\ref{ST2}) and the corresponding equations of motion for the polarization modes (\ref{eom}), we examined the presence of both the amplitude and velocity birefringence of the propagation of the gravitational waves in our model.
In particular, due to the presence of $\mathcal{W}_{-1}\frac{2a}{sk}$ term in the effective propagating speeds (\ref{cT2}), one of the two polarization modes of the GWs suffers from the gradient instability on large scales.

\acknowledgments

This work was partly supported by the National Natural Science Foundation of China (NSFC) under the grant No. 11975020 and No. 12005309.

\appendix

\section{Independent monomials}\label{app:indep}

When contracting with the Levi-Civita tensor, there are some identities which can be used to reduce the number of independent contractions. 
Consider an arbitrary tensor field carrying 3 indices $X_{\mu\nu\lambda}$, which we do not assume any symmetry a priori. Its completely antisymmetric part is defined by 
\begin{eqnarray}
	X_{[\rho\sigma\lambda]} & \coloneqq & \frac{1}{3!}\delta_{\rho\sigma\lambda}^{\rho'\sigma'\lambda'}X_{\rho'\sigma'\lambda'},
\end{eqnarray}
where $\delta_{\rho\sigma\lambda}^{\rho'\sigma'\lambda'}$ is the generalized Kronecker symbol.
On the one hand,
\begin{eqnarray}
	\epsilon^{\mu\nu\rho\sigma}X_{[\rho\sigma\lambda]} & = & \epsilon^{\mu\nu\rho\sigma}\frac{1}{6}\left(X_{\rho\sigma\lambda}-X_{\rho\lambda\sigma}+X_{\sigma\lambda\rho}-X_{\sigma\rho\lambda}+X_{\lambda\rho\sigma}-X_{\lambda\sigma\rho}\right)\nonumber \\
	& \equiv & \frac{1}{3}\epsilon^{\mu\nu\rho\sigma}\left(X_{\rho\sigma\lambda}+X_{\sigma\lambda\rho}+X_{\lambda\rho\sigma}\right).\label{epsX1}
\end{eqnarray}
On the other hand, since $\delta_{\rho\sigma\lambda}^{\rho'\sigma'\lambda'}\equiv-\epsilon_{\rho\sigma\lambda\alpha}\epsilon^{\rho'\sigma'\lambda'\alpha}$,
we have
\begin{eqnarray}
	\epsilon^{\mu\nu\rho\sigma}X_{[\rho\sigma\lambda]} & = & \epsilon^{\mu\nu\rho\sigma}\frac{1}{3!}\left(-\epsilon_{\rho\sigma\lambda\alpha}\epsilon^{\rho'\sigma'\lambda'\alpha}\right)X_{\rho'\sigma'\lambda'}=-\frac{1}{3!}\underbrace{\epsilon^{\mu\nu\rho\sigma}\epsilon_{\lambda\alpha\rho\sigma}}_{=-2!\delta_{\lambda\alpha}^{\mu\nu}}\epsilon^{\rho'\sigma'\lambda'\alpha}X_{\rho'\sigma'\lambda'}\nonumber \\
	& = & \frac{1}{3}\left(\delta_{\lambda}^{\mu}\epsilon^{\rho\sigma\lambda'\nu}-\delta_{\lambda}^{\nu}\epsilon^{\rho\sigma\lambda'\mu}\right)X_{\rho\sigma\lambda'}.\label{epsX2}
\end{eqnarray}
Comparing (\ref{epsX1}) and (\ref{epsX2}) yields the identity
\begin{equation}
	\left(\epsilon^{\mu\rho\sigma\alpha}\delta_{\lambda}^{\nu}-\epsilon^{\nu\rho\sigma\alpha}\delta_{\lambda}^{\mu}\right)X_{\rho\sigma\alpha}=\epsilon^{\mu\nu\rho\sigma}\left(X_{\rho\sigma\lambda}+X_{\sigma\lambda\rho}+X_{\lambda\rho\sigma}\right).\label{antisymm_id}
\end{equation}

For the $\epsilon QQ\phi\cdots\phi$ terms, naively we have 9 independent
contractions if only the (anti-)symmetries of $\epsilon$-tensor and
$Q_{\rho\mu\nu}$ are take into account, which we denote as\footnote{$\left[\epsilon QQ\right]_{1},\cdots , \left[\epsilon QQ\right]_{7}$ correspond to $L_{a1},\cdots,L_{a7}$ in \cite{Conroy:2019ibo}. Only 5 of them are independent.}
\begin{eqnarray*}
	\left[\epsilon QQ\right]_{1} & \coloneqq & \epsilon_{\mu\nu\rho\sigma}Q_{\phantom{\mu\nu}\lambda}^{\mu\nu}Q^{\rho\sigma\lambda}\equiv\left\{ \bm{1}\right\} ,\\
	\left[\epsilon QQ\right]_{2} & \coloneqq & \epsilon_{\mu\nu\rho\sigma}Q_{\phantom{\mu\nu}\alpha}^{\mu\nu}Q_{\phantom{\rho\sigma}\beta}^{\rho\sigma}\phi^{\alpha}\phi^{\beta}\equiv\left\{ \bm{1}\right\} _{\alpha\beta}\phi^{\alpha}\phi^{\beta},\\
	\left[\epsilon QQ\right]_{3} & \coloneqq & \epsilon_{\mu\nu\rho\beta}Q_{\phantom{\mu\nu}\lambda}^{\mu\nu}Q_{\alpha}^{\phantom{\alpha}\rho\lambda}\phi^{\alpha}\phi^{\beta}\equiv\left\{ \bm{2}\right\} _{\alpha\beta}\phi^{\alpha}\phi^{\beta},\\
	\left[\epsilon QQ\right]_{4} & \coloneqq & \epsilon_{\mu\nu\rho\beta}Q_{\phantom{\mu\nu}\lambda}^{\mu\nu}Q_{\phantom{\lambda\rho}\alpha}^{\lambda\rho}\phi^{\alpha}\phi^{\beta}\equiv\left\{ \bm{3}\right\} _{\alpha\beta}\phi^{\alpha}\phi^{\beta},\\
	\left[\epsilon QQ\right]_{5} & \coloneqq & \epsilon_{\mu\nu\rho\beta}Q_{\phantom{\mu\nu}\lambda}^{\mu\nu}Q_{\phantom{\rho\lambda}}^{\rho\lambda}{}_{\alpha}\phi^{\alpha}\phi^{\beta}\equiv\left\{ \bm{4}\right\} _{\alpha\beta}\phi^{\alpha}\phi^{\beta},\\
	\left[\epsilon QQ\right]_{6} & \coloneqq & \epsilon_{\mu\nu\rho\beta}Q_{\phantom{\mu\nu}\alpha}^{\mu\nu}Q_{\phantom{\rho}\sigma}^{\sigma\phantom{\rho}\rho}\phi^{\alpha}\phi^{\beta}\equiv\left\{ \bm{5}\right\} _{\alpha\beta}\phi^{\alpha}\phi^{\beta},\\
	\left[\epsilon QQ\right]_{7} & \coloneqq & \epsilon_{\mu\nu\rho\beta}Q_{\phantom{\mu\nu}\alpha}^{\mu\nu}Q_{\phantom{\rho\sigma}\sigma}^{\rho\sigma}\phi^{\alpha}\phi^{\beta}\equiv\left\{ \bm{6}\right\} _{\alpha\beta}\phi^{\alpha}\phi^{\beta},\\
	\left[\epsilon QQ\right]_{8} & \coloneqq & \epsilon_{\mu\nu\lambda\beta}Q_{\phantom{\mu\nu}\alpha}^{\mu\nu}Q_{\rho\sigma}^{\phantom{\rho\sigma}\lambda}\phi^{\alpha}\phi^{\beta}\phi^{\rho}\phi^{\sigma}\equiv\left\{ \bm{1}\right\} _{\alpha\beta\rho\sigma}\phi^{\alpha}\phi^{\beta}\phi^{\rho}\phi^{\sigma},\\
	\left[\epsilon QQ\right]_{9} & \coloneqq & \epsilon_{\mu\nu\lambda\beta}Q_{\phantom{\mu\nu}\alpha}^{\mu\nu}Q_{\phantom{\lambda}\rho\sigma}^{\lambda}\phi^{\alpha}\phi^{\beta}\phi^{\rho}\phi^{\sigma}\equiv\left\{ \bm{2}\right\} _{\alpha\beta\rho\sigma}\phi^{\alpha}\phi^{\beta}\phi^{\rho}\phi^{\sigma}.
\end{eqnarray*}
By employing the identity (\ref{antisymm_id}), we find three relations
\begin{equation}
	\left\{ \bm{3}\right\} _{\alpha\beta}-\left\{ \bm{4}\right\} _{\alpha\beta}=-\left\{ \bm{1}\right\} _{\alpha\beta}-\left\{ \bm{5}\right\} _{\alpha\beta}+\left\{ \bm{6}\right\} _{\alpha\beta},\label{QQcontra_rel_1}
\end{equation}
\begin{equation}
	-\left\{ \bm{2}\right\} _{\alpha\beta}+\left\{ \bm{4}\right\} _{\alpha\beta}=\frac{1}{2}g_{\alpha\beta}\left\{ \bm{1}\right\} ,\label{QQcontra_rel_2}
\end{equation}
and
\begin{equation}
	-\left\{ \bm{1}\right\} _{\mu\nu\beta\alpha}+\left\{ \bm{2}\right\} _{\mu\nu\alpha\beta}\simeq\frac{1}{2}g_{\beta\nu}\left\{ \bm{1}\right\} _{\mu\alpha}.\label{QQcontra_rel_3}
\end{equation}
where ``$\simeq$'' means ``effective'' equivalence when contracting
with $\phi^{\alpha}\phi^{\beta}\phi^{\mu}\phi^{\nu}$ (i.e.,
when all the indices are completely symmetrized). As a result, out
of the 9 contractions, only 6 of them are independent, which we choose
to be $\mathcal{F}_{1},\cdots,\mathcal{F}_{6}$ in (\ref{calF1})-(\ref{calF6}).

\end{document}